\documentclass[aps,prl,twocolumn,reprint,amsmath,amssymb,superscriptaddress,floatfix,footinbib,longbibliography]{revtex4-2}

\usepackage{epsfig}
\usepackage{graphicx}
\usepackage{epstopdf}
\usepackage[T1]{fontenc}
\usepackage[applemac]{inputenc}
\usepackage{lmodern}
\usepackage[english]{babel}
\usepackage{lipsum}
\usepackage{ae}
\usepackage{units}

\usepackage{amsmath,amssymb,natbib,bm}
\usepackage{psfrag}
\usepackage{subfigure}
\usepackage{amsthm}
\usepackage{bbm}
\usepackage{booktabs}

\usepackage{tikz}
\usetikzlibrary{arrows}

\usepackage{color}
\usepackage{url}
\usepackage[colorlinks]{hyperref}
\hypersetup{%
        plainpages=true,
        breaklinks=true,
        hypertexnames=false,
        pageanchor=true,
        colorlinks=true,
        linkcolor={blue},
        citecolor={magenta},
        urlcolor={blue},
        anchorcolor={black}
      }
\usepackage[all]{hypcap} 

\usepackage{mleftright} 

\renewcommand{\eqref}[1]{\mbox{Eq.~(\ref{#1})}}

\newcommand{\figpanel}[2]{Fig.~\hyperref[#1]{\ref*{#1}(#2)}}
\newcommand{\figpanels}[3]{Fig.~\hyperref[#1]{\ref*{#1}(#2)-(#3)}}
\newcommand{\figpanelNoPrefix}[2]{\hyperref[#1]{\ref*{#1}(#2)}}

\newcommand{\brakket}[3]{\mleft\langle #1\mleft| #2 \mright| #3\mright\rangle}
\newcommand{\expec}[1]{\mleft\langle #1 \mright\rangle}

\newcommand{\comm}[2]{\mleft[ #1, #2 \mright]}

\newcommand{\be}{\begin{equation}}
\newcommand{\ee}{\end{equation}}
\newcommand{\bea}{\begin{eqnarray}}
\newcommand{\eea}{\end{eqnarray}}

\DeclareMathOperator{\Tr}{Tr} 

\newcommand{\ket}[1]{\vert #1 \rangle}
\newcommand{\bra}[1]{\langle #1 \vert}

\newcommand{\proj}[1]{\vert #1 \rangle \langle #1 \vert}

\begin{document}

\title{Universal fidelity reduction of quantum operations from weak dissipation}
\date{\today}

\author{Tahereh Abad}
\email{Tahereh.Abad@chalmers.se}
\affiliation{Department of Microtechnology and Nanoscience, Chalmers University of Technology, 412 96 Gothenburg, Sweden}

\author{Jorge Fern\'andez-Pend\'as}
\affiliation{Department of Microtechnology and Nanoscience, Chalmers University of Technology, 412 96 Gothenburg, Sweden}

\author{Anton Frisk Kockum}
\affiliation{Department of Microtechnology and Nanoscience, Chalmers University of Technology, 412 96 Gothenburg, Sweden}

\author{G\"oran Johansson}
\email{Goran.L.Johansson@chalmers.se}
\affiliation{Department of Microtechnology and Nanoscience, Chalmers University of Technology, 412 96 Gothenburg, Sweden}

\begin{abstract}

Quantum information processing is in real systems often limited by dissipation, stemming from remaining uncontrolled interaction with microscopic degrees of freedom. Given recent experimental progress, we consider weak dissipation, resulting in a small error probability per operation. Here, we find a simple formula for the fidelity reduction of any desired quantum operation. Interestingly, this reduction is independent of the specific operation; it depends only on the operation time and the dissipation. Using our formula, we investigate the situation where dissipation in different parts of the system have correlations, which is detrimental for the successful application of quantum error correction. Surprisingly, we find that a large class of correlations gives the same fidelity reduction as uncorrelated dissipation of similar strength.

\end{abstract}

\maketitle

\paragraph*{Introduction.}

Numerous architectures are being explored for quantum computers~\cite{Ladd2010}, e.g., circuit quantum electrodynamics~\cite{Wendin2017, Gu2017, Krantz2019, Blais2021}, trapped ions~\cite{Haffner2008, Bruzewicz2019}, quantum dots~\cite{Chatterjee2021}, and photonics~\cite{Flamini2019}. The long-term goal is solving useful problems where classical computers fall short~\cite{Montanaro2016} and in the nearer term to outperform classical supercomputers for specific computing tasks~\cite{Arute2019}. However, uncontrolled interactions between the quantum system and its surroundings destroy quantum coherence and thus reduce the fidelity of the quantum operations (gates). How to create high-fidelity quantum gates in the presence of this environmentally induced decoherence is probably the most important problem to solve, both for near-term quantum computation and for the long-term goal of fault-tolerant quantum computing~\cite{Preskill1998, Knill2010, Ladd2010}. 

With gate fidelities approaching the fault-tolerant threshold, characterizing and reducing the remaining errors becomes increasingly challenging~\cite{Benhelm2008, Barends2014}. A well-used tool for characterization is Clifford-based randomized benchmarking (RB)~\cite{Knill2008, Magesan2011}, which enables mapping gate errors onto control parameters and feeding this back to optimize the gates. With optimized control, the fidelity is limited by decoherence processes such as energy decay and dephasing. Explicit analytical expressions for this fidelity reduction has been derived for single-qubit Clifford gates~\cite{OMalley2015} as well as certain two-qubit gates~\cite{Sete2021, Arute2019}, to first order in the ratio between the gate time $\tau$ and the decoherence time $1 / \Gamma$.

Here, we derive general analytical results showing how the fidelity of single-, two-, as well as general multi-qubit gates is affected by weak decoherence. We consider the standard model for interaction with a Markovian environment, using a Lindblad master equation, and find that to first order in $\Gamma \tau$, the reduction in fidelity is \textit{independent of the specific gate}. This result holds for all single- and multi-qubit gates where the evolution is confined to the computational subspace. It also holds for the case when different qubits see the same environment, i.e., correlated multi-qubit noise processes. We discuss in particular the effect of energy relaxation and dephasing, and give explicit formulas for the reduction of the average gate fidelity, which only depends on the number of qubits involved in the gate and the rates of the decoherence processes affecting those qubits. We then explicitly explore the difference between uncorrelated and two scenarios of fully correlated multi-qubit dissipation. Our results provide bounds that allow for robust estimation and optimization of single- as well as multi-qubit gate fidelities, and may enable establishing constraints on the power of noisy quantum computers. 

\paragraph*{Method.}

The average gate fidelity $\overline{F}$ of a trace-preserving quantum operation $\mathcal{E}$, acting on an $N$-qubit system, is defined as
\be \label{fidelity_integral}
 \overline{F} \equiv \int d\psi \brakket{\psi}{U_g^\dag \mathcal{E}(\psi) U_g}{\psi},
\ee
where the integral is over all pure initial states and $U_g$ is the unitary operator corresponding to the ideal gate operation. Note that $\overline{F} = 1$ if and only if ${\cal E}$ implements $U_g$ perfectly, while lower values indicate that ${\cal E}$ is a noisy or otherwise imperfect implementation of $U_g$. The gate operation in \eqref{fidelity_integral} can be generated by a time-dependent Hamiltonian $H(t)$ applied for a time $\tau$, such that $U_g=U(0,\tau)$, where we have introduced the time-evolution operator for the ideal gate operation $U(t_1,t_2) = {\cal T} \exp[-\frac{i}{\hbar}\int_{t_1}^{t_2} H(t) dt]$ and ${\cal T}$ is the time-ordering operator.

We describe the effect of decoherence using the standard Lindblad superoperator
\bea
\mathcal{D}[\hat L] \rho = \hat L \rho \hat L^\dagger - \frac{1}{2} \mleft\{\hat L^\dagger \hat L, \rho \mright\}
\eea
acting on the system density matrix $\rho$. The time evolution of the system with $N_L$ dissipative processes is then given by the master equation
\begin{equation} \label{master}
\dot\rho(t) = -\frac{i}{\hbar} \comm{H(t)}{\rho(t)} + \sum_{k = 1}^{N_L} \Gamma_k \mathcal{D} [\hat{L}_k] \rho(t),
\end{equation}
where each process has its corresponding rate $\Gamma_k$ and Lindblad jump operator $\hat{L}_k$. We will later discuss specifically energy relaxation and dephasing acting on individual qubits, but for now the jump operators can be any multi-qubit operator. Inspired by the current experimental state-of-the-art, where incoherent errors are on the percent level or less~\cite{Bengtsson2020, Sung2021, Negirneac2021, Stehlik2021, Srinivas2021, Clark2021}, we now expand the solution to the master equation in the small parameter $\Gamma_{k} \tau \ll 1$ for a pure initial state $\ket{\psi}$. The unperturbed solution is simply $\rho_\psi^{(0)}(t)=\proj{\psi(t)}$, where $\ket{\psi(t)}=U(0,t)\ket{\psi}$. The first-order correction due to the $k$th decoherence process is~\cite{Villegas-Martinez2016}
\be \label{psi_first_order}
\rho_{\psi,k}^{(1)}(t) = \Gamma_{k}\int_0^t dt' U(t',t) \mleft[ \mathcal{D}[\hat{L}_k]\rho_\psi^{(0)}(t')\mright]U(t',t)^\dag,
\ee
which corresponds to applying the dissipator ${D}[\hat{L}_k]$ to the ideal pure state $\ket{\psi(t')}$ once, at any time $t'<t$.

\paragraph*{Main result.}

Each dissipative process contributes independently to first order, and with this correction to the ideal density-matrix at the end of the gate, we can evaluate the gate fidelity using \eqref{fidelity_integral}:
\be \label{delta_F_diss}
 \overline{F} = 1 + \sum_{k = 1}^{N_L}\int d\psi  \bra{\psi} U(0,\tau)^\dag \rho_{\psi,k}^{(1)}(\tau) U(0,\tau) \ket{\psi}.
\ee
Inserting \eqref{psi_first_order} and first performing the integral over initial states $\int d\psi$, we find  
\bea
\label{deltaF}
\int d\psi \mleft[ \bra{\psi(t')} \hat{L}\proj{\psi(t')}\hat{L}^\dag \ket{\psi(t')} 
 -\bra{\psi(t')} \hat{L}^\dagger \hat{L} \ket{\psi(t')} \mright] \notag \\
= \int d\psi \mleft[\bra{\psi} \hat{L} \proj{\psi} \hat{L}^\dag \ket{\psi} -\bra{\psi} \hat{L}^\dag \hat{L} \ket{\psi} \mright] \equiv \delta F (\hat{L}). \qquad\:\:
\eea
The first expression contains only expectation values of jump operators with respect to the intermediate pure state $\ket{\psi(t')}$. Since the unitary gate evolution $U(0,t')$ only performs a rotation in Hilbert space, it leaves the set of {\em all} initial states $\ket{\psi}$ invariant. Integrating over all initial states $\ket{\psi}$ in \eqref{delta_F_diss} is thus identical to integrating over all states $\ket{\psi(t')}$ for {\em any} $t'$. This renders the remaining integrand {\em time-independent} such that from the remaining time integral we trivially obtain
\be \label{fidelity_first_order}
\bar{F} = 1 + \tau \sum_{k= 1}^{N_L} \Gamma_k\, \delta F(\hat{L}_k) + \mathcal{O}\mleft(\tau^2 \Gamma_k^2 \mright).
\ee
This is \textit{the main result} of this article. The reduction of gate fidelity is thus {\em independent} of which unitary gate $U_g$ is performed and proportional to the time $\tau$ it takes to perform the gate. Each dissipative channel contributes independently, proportional to its rate $\Gamma_k$ and the factor $\delta F(\hat{L}_k)$, which we now proceed to evaluate for a few relevant processes.

\paragraph*{General formula for fidelity reduction of $N$-qubit gates.}

To evaluate the integral over all pure states in \eqref{deltaF}, we first rewrite it using a density-matrix representation,
\be
\label{deltaFdensity}
\delta F (\hat{L}) = \int d\psi \mleft( \Tr{\mleft[\hat{L}^\dag \rho_\psi \hat{L} \rho_\psi \mright]} - \Tr{\mleft[ \hat{L}^\dag \hat{L} \rho_\psi \mright]} \mright). 
\ee
In the case of a single qubit, we can expand the density-matrix in four terms: $\rho_\psi = \frac{1}{2} \mleft( \sigma_0 + n_x\sigma_x + n_y\sigma_y + n_z\sigma_z \mright)$, where $\sigma_0$ is the $2 \times 2$ identity matrix and $\sigma_i$ for $i\in\{x,y,z\}$ are the corresponding Pauli matrices. Inserting this expression in \eqref{deltaFdensity}, the first term expands into 16 terms, while the second gives four terms. The average over $d\psi$ now corresponds to a an integral over the three real-valued coefficients $n_x$, $n_y$, and $n_z$ under the normalisation constraint $n_x^2+n_y^2+n_z^2=1$, i.e., the Bloch sphere. This can be calculated explicitly~\cite{Bowdrey2002}, but here we follow Ref.~\cite{Cabrera2007} and note that the symmetries of the Hilbert space imply that $\langle n_i \rangle = 0$ and $\langle n_i n_j \rangle = \delta_{ij}/3$ for $i,j \in \{x,y,z\}$, where angular brackets denote integration over the Bloch sphere, $\delta_{ij}$ is the Kronecker delta, and the factor $1/3$ follows from the normalisation.
Thus, for a single qubit \eqref{deltaFdensity} reduces to
\be \label{dF_single_1}
\delta F_1(\hat{L})= - \frac{1}{4}\Tr{\left[
\hat{L}^\dagger \hat{L} \right]}+\frac{1}{12} \sum_{j\in\{x,y,z\}} \Tr\left[
\hat{L}^\dagger \sigma_j \hat{L} \sigma_j  \right].
\ee

For a system with $N$ qubits, we can expand any density-matrix in a basis consisting of all $4^N$ possible tensor-product combinations of Pauli matrices and identity. The element consisting of only identity matrices is the identity matrix in $d=2^N$ dimensions and thus has trace $d$, fixing the overall normalisation to $1/d$. Denoting the other $d^2-1$ traceless basis matrices as $\hat{f}_i$, we write
$\rho_\psi = \frac{1}{d} \mleft( \hat{1}_d + \sum_{i=1}^{d^2-1} n_i \hat{f}_i \mright)$. We find similar rules for averages over the real-valued coefficients $n_i$ as in the single-qubit case~\cite{Cabrera2007, SuppMat}: $\langle n_i \rangle = 0$ and $\langle n_i n_j \rangle = \delta_{ij}/(d+1)$. Thus, for operations on $N$ qubits, \eqref{deltaFdensity} reduces to
\be \label{dF_multi}
\delta F_N(\hat{L}) = \frac{1-d}{d^2} \Tr{\mleft[ \hat{L}^\dag \hat{L} \mright]} + \frac{\sum_{i=1}^{d^2-1} \Tr\mleft[ \hat{L}^\dag \hat{f}_i \hat{L} \hat{f}_i  \mright]}{d^2(d+1)} ,
\ee
giving a general formula for the reduction of fidelity of general $N$-qubit gates to first order in Markovian dissipation. The expression is indeed gate-independent, but depends on the nature of the dissipative processes, expressed through the corresponding Lindblad jump operator $\hat{L}$. We now proceed to discuss different forms of this operator, in particular the difference between processes that act independently or in a correlated fashion on different qubits.

\paragraph*{Effect of uncorrelated relaxation and dephasing.}

We first consider individual qubit energy relaxation acting on one qubit with jump operator $\hat{L}=\sigma_-$ and rate $\Gamma_1$, and additional pure dephasing with jump operator $\sigma_z$ and rate $\Gamma_\phi$ [note that the rate multiplying the dissipator in \eqref{master} is $\Gamma_\phi/2$, making the coherences decay with the rate $\Gamma_\phi$]. For uncorrelated dissipation, the $N$-qubit jump operators are tensor products with identity matrices acting on all other qubits. Since the trace operations in \eqref{dF_multi} then factorise into products of single-qubit traces, we straightforwardly find
\be
\delta F_N (\sigma_z^1 \otimes \sigma_0^2 \dots \sigma_0^N) = 2 \delta F_N (\sigma_-^1 \otimes \sigma_0^2 \dots \sigma_0^N) = -\frac{d}{d+1},
\ee
extending the expressions for single- and two-qubit Clifford gates given in Ref.~\cite{Arute2019} to arbitrary gates on an arbitrary number of qubits.

Remembering that different dissipators add independently to the gate fidelity according to \eqref{fidelity_first_order}, we can then find the first-order reduction in gate fidelity due to uncorrelated energy relaxation and pure dephasing on all $N$ qubits~\cite{SuppMat}:
\be \label{fidelity_first_order_uncorrelated}
\bar{F}^{uc}_N = 1  -\frac{d}{2(d+1)} \tau \sum_{k=1}^N \mleft(\Gamma_1^k+\Gamma_\phi^k \mright),
\ee
where $\Gamma_{1/\phi}^k = 1 / T_{1 / \phi}^k$ is the relaxation/dephasing rate of qubit $k$ and $T_{1 / \phi}^k$ the relaxation/dephasing time.
By comparing the single and two-qubit gate fidelities in an experimental system, this formula allows to assess to what extent the gates are decoherence-limited. However, as we will see below, a multi-qubit gate error that agrees with this expression does not guarantee that the noise processes are indeed uncorrelated between qubits.

We stress that the gate-independence of the gate fidelity is only valid to first order in the dissipative correction. For single-qubit rotations around the $x$ and $z$ axes, this is illustrated by the analytical solutions to the master equation with energy relaxation and dephasing, which, for $\pi$ rotations, to second order in the dissipation yield the gate fidelities
\begin{align} \label{}
\bar F_{\sigma_x} &= 1 - \frac{\Gamma_1 + \Gamma_\phi}{3} \tau + \frac{1}{8} \mleft( \frac{11}{12} \Gamma_1^2 + \frac{5}{3} \Gamma_1 \Gamma_\phi + \Gamma_\phi^2 \mright )\tau^2, \\
\bar F_{\sigma_z} &= 1 - \frac{\Gamma_1 + \Gamma_\phi}{3} \tau + \frac{1}{8} \mleft( \Gamma_1^2 + \frac{4}{3} \mleft[ \Gamma_1 \Gamma_\phi + \Gamma_\phi^2 \mright] \mright) \tau^2.
\end{align}
The limitation of gate independence to first order is also clear from the fact that in the second-order expansion, the dissipator will act two times, so the expression include averages which are not over the full Hilbert space and thus depend on the relation between the gate operation and the dissipation.

\paragraph*{Results for correlated noise.}

Dissipation can be correlated in many different ways. Here, we discuss two cases, which affect the gate fidelity differently. First, we treat correlated decoherence arising from many qubits connected to the same environmental mode. For simplicity, we consider equal coupling of all $N$ qubits, leading to the jump operator $\hat{L}^N_{\phi c} = \sum_{k=1}^N \sigma_z^k$ describing correlated dephasing with rate $\Gamma_{\phi c}$ [again corresponding to a rate $\Gamma_{\phi c}/2$ in \eqref{master}], as well as the jump operator $\hat{L}^N_{1c} = \sum_{k=1}^N \sigma_-^k$ describing correlated relaxation with rate $\Gamma_{1c}$. The correlated dephasing corresponds to a decay of the coherence between two multi-qubit states, with a rate $\delta n^2 \Gamma_{\phi c}$, where $\delta n$ is the difference in excitation number between the two states~\cite{Jeske2013}. In a three-qubit system, the coherence between $\ket{000}$ and $\ket{111}$ thus decays with the rate $9\Gamma_{\phi c}$, while the subspace spanned by states with the same number of excitations, e.g., $\ket{100}$, $\ket{010}$, and $\ket{001}$, is not affected by dephasing. In a similar fashion, the correlated relaxation gives rise to non-decaying multi-qubit dark states as well as bright states decaying quickly due to superradiance. 

We can straightforwardly evaluate the reduction of $N$-qubit gate fidelity due to correlated dephasing and relaxation using \eqref{dF_multi}, finding~\cite{SuppMat}
\be
\bar{F}^c_N = 1 - \frac{N d}{2(d+1)} \tau \mleft( \Gamma_{1c} + \Gamma_{\phi c} \mright),
\label{eq:dF_Corr_Deph_Rel}
\ee
which somewhat surprisingly is identical to the reduction in fidelity when all $N$ qubits are subject to uncorrelated dephasing with rate $\Gamma_{\phi c}$ and uncorrelated relaxation with rate $\Gamma_{1c}$. This illustrates that the average gate fidelity is \textit{not} a sensitive probe for detecting whether the dissipation arises from this type of additive linear coupling to a common bath. Averaging over all initial states tends to hide the fact that this type of correlated dissipation acts very differently on different parts of the computational Hilbert space and thus creates correlated errors between qubits, which is potentially detrimental for quantum error correction~\cite{Wilen2021}.

Finally, consider instead a two-photon relaxation process, where two qubits can relax to a bath accepting only the sum of the two qubit energies, corresponding to the jump operator $\hat{L}_{2p} = \sigma_- \otimes \sigma_-$ and a rate $\Gamma_{2p}$. If one measures the relaxation time of the qubits individually, with the other qubits in their ground states, this process will not contribute. However, for the two-qubit gate fidelity one finds an extra reduction~\cite{SuppMat},
\be
\bar{F} = \bar{F}^{uc}_2 - \frac{\Gamma_{2p}\tau}{5},
\label{eq:dF_TwoPhoton}
\ee
which would add to the reduction predicted by the measured single-qubit relaxation and dephasing rates. The average two-qubit gate fidelity can thus detect this type of two-photon relaxation processes.






\begin{acknowledgments}

\paragraph*{Acknowledgements.}

We acknowledge useful discussions with Andreas Bengtsson and John Martinis. 

We acknowledge support from the Knut and Alice Wallenberg Foundation through the Wallenberg Centre for Quantum Technology (WACQT) and from the EU Flagship on Quantum Technology H2020-FETFLAG-2018-03 project 820363 OpenSuperQ.

\end{acknowledgments}

\bibliography{ReferencesFidelity}

\onecolumngrid
\clearpage
\setcounter{section}{0}
\renewcommand{\thesection}{S\arabic{section}}
\setcounter{equation}{0}
\renewcommand{\theequation}{S\arabic{equation}}
\renewcommand{\thetable}{S\arabic{table}}
\renewcommand{\bibnumfmt}[1]{[S#1]}
\renewcommand{\citenumfont}[1]{S#1}

\setcounter{page}{1}


\section{Fidelity correction for 2-qubit gates}

In this appendix, we show in more detail how to calculate the fidelity correction for $N$-qubit gates from \eqref{deltaF} in the main text to obtain the expression for $\delta F_N(\hat{L})$ given in \eqref{dF_multi} in the main text. For clarity, we first show how the result obtained in the single-qubit case [\eqref{dF_single_1} in the main text] is extended to two qubits, and how that expression simplifies when dissipation just acts on one of the two qubits, before tackling the $N$-qubit case.

The key is to define an appropriate basis for the density matrix. For two qubits, the density matrix can be expanded in terms of tensor products of Pauli matrices as
\be \label{rho2}
\rho = \frac{1}{4} \mleft( \sigma_0^1 \sigma_0^2 + \sum_{i,j \neq 0,0} c_{ij} \sigma_i^1 \sigma_j^2 \mright),
\ee
where the coefficient in front of the only non-traceless matrix, the identity matrix $\sigma_0^1 \otimes \sigma_0^2$, is determined by the condition $\Tr{\rho} = 1$. For any pure state, we then obtain from $\Tr{\rho^2}=1$ that
\be
\frac{1}{4} \mleft( 1 + \sum_{i,j \neq 0,0} \mleft( c_{ij} \mright)^2 \mright) = 1,
\label{eq:TracePureStateSquared}
\ee
which gives $\langle \sum_{ij} \mleft( c_{ij} \mright)^2 \rangle = 3$. By symmetry, the contribution from each of the 15 components (excluding $i=j=0$) must be the same, i.e,
\be
\expec{\sum_{i,j \neq 0,0} \mleft( c_{ij} \mright)^2} = 15 \expec{\mleft( c_{ij} \mright)^2}. 
\ee
So $\expec{\mleft( c_{ij} \mright)^2} = 1/5$, and more generally
\be
\expec{c_{ij} c_{kl}} = \delta_{ik} \delta_{jl} \expec{\mleft( c_{ij} \mright)^2} = \frac{1}{5} \delta_{ik} \delta_{jl}.
\label{eq:ExpecCijCkl}
\ee
Since there are also unitary transformations that simply change the sign of the coefficients $c_{ij}$, the average $\rho$ is unpolarized: $\expec{c_{ij}} = 0$.

Inserting \eqref{rho2} into \eqref{deltaF} from the main text gives
\begin{align} \label{df_2Q}
\delta F_2 (\hat{L}) =\ & \frac{1}{16} \int d \psi \text{Tr} \mleft[ \hat{L} \mleft( \sigma_0^1 \sigma_0^2 + \sum_{i,j \neq 0,0} c_{ij} \sigma_i^1 \sigma_j^2 \mright) \hat{L}^\dag \mleft( \sigma_0^1 \sigma_0^2 + \sum_{k,l \neq 0,0} c_{kl} \sigma_k^1 \sigma_l^2 \mright) \mright] \nonumber\\
& - \frac{1}{4} \int d \psi \text{Tr} \mleft[ \hat{L}^\dag \hat{L} \mleft(  \sigma_0^1 \sigma_0^2 + \sum_{i,j \neq 0,0} c_{ij} \sigma_i^1 \sigma_j^2 \mright) \mright].
\end{align}
Using the relations in Eqs.~(\ref{eq:TracePureStateSquared})--(\ref{eq:ExpecCijCkl}), averaging over all possible initial states $\ket{\psi}$ reduces \eqref{df_2Q} to
\be
\delta F_2 (\hat{L}) = - \frac{3}{16} \text{Tr} \mleft[ \hat{L} \hat{L}^\dag \mright] + \frac{1}{80} \sum_{i,j \neq 0,0} \text{Tr} \mleft[ \hat{L} \mleft( \sigma_i^1 \sigma_j^2 \mright) \hat{L}^\dag \mleft( \sigma_i^1 \sigma_j^2 \mright) \mright] ,
\ee
where the first term is the sum of the last term in \eqref{df_2Q} and the term $\hat{L} \sigma_0^1 \sigma_0^2 \hat{L}^\dag \sigma_0^1 \sigma_0^2 = \hat{L} \hat{L}^\dag$ from the first term in \eqref{df_2Q}. 
For jump operators on the simple form $\hat{L} = \hat{L}_1 \otimes \sigma_0^2$, i.e., uncorrelated dissipation acting on qubit 1 alone, we obtain
\begin{align}
\delta F_2 (\hat{L}_1 \otimes \sigma_0^2) &= \mleft( - \frac{3 \times 2}{16} + \frac{3 \times 2}{80} \mright) \text{Tr} \mleft[ \hat{L}_1 \hat{L}_1^\dag \mright] + \frac{2}{80} \sum_{i=1}^3 \text{Tr} \mleft[ \hat{L}_1 \sigma_i \hat{L}_1^\dag \sigma_i \mright] \nonumber\\
&= -\frac{3}{10} \text{Tr} \mleft[ \hat{L}_1 \hat{L}_1^\dag \mright] + \frac{1}{40} \sum_{i=1}^3 \text{Tr} \mleft[
\hat{L}_1 \sigma_i \hat{L}^\dag \sigma_i \mright] ,
\end{align}
where the three terms with $i = 0$ and $j \neq 0$ from the double sum contribute to the term proportional to $\Tr{(\hat{L}_1 \hat{L}_1^\dag)}$.


\section{$N$-qubit gates}

For the general $N$-qubit case, the basis elements for the density matrix are products of $N$ Pauli matrices $\sigma_{\mu}$. The basis, excluding the identity element, contains $d^2 - 1$ elements, where $d = 2^N$, and the density matrix can be expanded as
\be \label{rhoN}
\rho = \frac{1}{d} \mleft( \sigma_0^1 \ldots \sigma_0^N + \sum_{i_1, \ldots, i_N} c_{i_1 \ldots i_N} \sigma_{i_1}^1 \ldots \sigma_{i_N}^N \mright) \equiv \frac{1}{d} \mleft( \hat{1}_d + \sum_{i = 1}^{d^2 - 1} c_i \hat{f}_i \mright) ,
\ee
where the term $i_1 = i_2 = \dots = i_N = 0$ is taken outside the sum. Here, we defined the set of basis matrices $\hat{f}_i$, consisting of the $d^2-1$ tensor products of Pauli matrices, and then collected the $N$ indices $i_1 \ldots i_N$ into the single combined index $1\leq i \leq d^2-1$.

For any pure state, we obtain from $\Tr{\rho^2} = 1$ that
\be
\frac{1}{d} \mleft( 1+\sum_i^{d^2-1} c_i^2 \mright) = 1,
\ee
which implies $\expec{\sum_i c_i^2} = d - 1$. Since, as in the two-qubit case, symmetry gives that the average of each component contributes the same amount, we have
\be
\expec{c_i^2} = \frac{d - 1}{d^2 - 1} = \frac{1}{1 + d},
\ee
and, more generally,
\be
\expec{\mleft( c_{i_1, \ldots, i_N} \mright) \mleft( c_{j_1, \ldots, j_N} \mright)} = \frac{1}{1+d} \delta_{i_1 j_1} \ldots \delta_{i_N j_N} \Leftrightarrow \expec{c_i c_j} = \frac{\delta_{ij}}{d+1}.
\ee
Inserting \eqref{rhoN} into \eqref{deltaF} from the main text and averaging over all possible initial states $\ket{\psi}$, using these relations, the fidelity reduction for the $N$-qubit case becomes
\be
\label{deltaFmulti}
\delta F_N (\hat{L}) = \frac{1-d}{d^2} \Tr{\mleft[ \hat{L}^\dag \hat{L} \mright]} + \frac{1}{d^2(d+1)} \sum_{i = 1}^{d^2 - 1} \Tr{\mleft[ \hat{L} \hat{f}_i \hat{L}^\dag \hat{f}_i \mright]} ,
\ee
which is \eqref{dF_multi} in the main text.


\section{Uncorrelated noise}

Here we present the details of the derivation of the result in \eqref{fidelity_first_order_uncorrelated} in the main text, which quantifies the fidelity reduction in an $N$-qubit system from uncorrelated energy relaxation and pure dephasing. For a single qubit, energy relaxation is described by the jump operator $\hat{L}=\sigma_-$ and the relaxation rate $\Gamma_1  = 1 / T_1$. Pure dephasing is described by the jump operator $\hat{L}=\sigma_z$ and the pure dephasing rate $\Gamma_\phi = 1 / T_\phi$ [note that the rate multiplying the dissipator in \eqref{master} in the main text is $\Gamma_\phi/2$, making the coherences decay with the rate $\Gamma_\phi$]. In an $N$-qubit system, uncorrelated energy relaxation acting on qubit number $k$ is described by a jump operator consisting of $\sigma_-^k$ acting on qubit $k$ in a tensor product with identity matrices acting on all other $N-1$ qubits. We denote the corresponding relaxation rate with $\Gamma_1^k$. Uncorrelated pure dephasing is modeled in a similar fashion with non-trivial jump operator $\sigma_z^k$ and rate $\Gamma_\phi^k$.

In evaluating the gate fidelity reduction, it is useful to note that
\be
\label{cancellation}
\sum_{i=0}^3 \Tr{\mleft[ \sigma_- \sigma_i \sigma_+ \sigma_i \mright]} =\sum_{i=0}^3 \Tr{\mleft[\sigma_z \sigma_i \sigma_z \sigma_i\mright]} =0.
\ee
To use this identity, we rewrite \eqref{deltaFmulti} by extending the summation in the second term to also include $i = 0$, i.e., the $N$-qubit identity matrix $\hat{f}_0=\sigma_0^1 \otimes \sigma_0^2 \otimes \dots \otimes \sigma_0^N$, and subtracting the corresponding quantity from the first term, yielding
\be
\label{deltaFmulti2}
\delta F_N(\hat{L})= -\frac{1}{d+1} \Tr{\mleft[\hat{L}^\dag \hat{L}\mright]} 
+\frac{1}{d^2(d+1)} 
\sum_{i=0}^{d^2-1}\Tr{\mleft[ \hat{L} \hat{f}_i \hat{L}^\dag \hat{f}_i\mright]}.
\ee
Considering energy relaxation on qubit 1, we find, by noticing that the tensor-product form of the jump operator makes the trace take the form of a product of the traces over the individual qubits,
\bea
\label{deltaFrel}
\delta F_N(\sigma_-^1\otimes \sigma_0^2 \dots \otimes \sigma_0^N) &=& -\frac{1}{d+1} \Tr{\mleft[\sigma_+\sigma_- \mright]}\left(\Tr{\mleft[\sigma_0 \mright]}\right)^{N-1} 
+\frac{1}{d^2(d+1)} 
\sum_{i=0}^{3}\Tr{\mleft[ \sigma_- \sigma_i \sigma_+ \sigma_i\mright]}\left(\Tr{\mleft[\sigma_0\mright]}\right)^{N-1} \notag \\
&=& -\frac{1}{d+1} (1 \times 2^{N-1}) +\frac{1}{d^2(d+1)} 
(0 \times 2^{N-1}) = -\frac{d}{2(d+1)}.
\eea
In a similar fashion, we find for pure dephasing
\bea
\label{deltaFdeph}
\delta F_N(\sigma_z^1\otimes \sigma_0^2 \dots \otimes \sigma_0^N) &=& -\frac{1}{d+1} \Tr{\mleft[\sigma_z\sigma_z\mright]}\mleft(\Tr{\mleft[\sigma_0\mright]}\mright)^{N-1} 
+\frac{1}{d^2(d+1)} 
\sum_{i=0}^{3}\Tr{\mleft[\sigma_z \sigma_i \sigma_z \sigma_i\mright]}\mleft(\Tr{\mleft[\sigma_0\mright]}\mright)^{N-1} \notag \\
&=& -\frac{1}{d+1} (2 \times 2^{N-1}) +\frac{1}{d^2(d+1)} 
(0 \times 2^{N-1}) = -\frac{d}{d+1}.
\eea
Remembering that different dissipators add incoherently to the gate fidelity according to \eqref{fidelity_first_order}, it is straightforward to write down the first-order reduction in gate fidelity due to uncorrelated energy relaxation and pure dephasing:
\be
\bar{F} = 1  -\frac{d}{2(d+1)} \tau \sum_{k=1}^{N} \mleft(\Gamma_1^k+\Gamma_\phi^k \mright),
\ee
which is \eqref{fidelity_first_order_uncorrelated} in the main text.


\section{Correlated noise}

Here we give the details for calculating fidelity reduction due to correlated relaxation and dephasing in an $N$-qubit system. As a warm-up, we first consider the case of two qubits. Inserting the jump operator for two qubits coupled to the same low-frequency bath inducing correlated dephasing, $\hat{L}=\sigma_z^1+\sigma_z^2$, into \eqref{deltaFmulti2}, yields a fidelity reduction
\bea
\label{deltaF2corrDeph}
\delta F_2(\sigma_z^1+\sigma_z^2)&=& -\frac{1}{4+1} \Tr{\mleft[2\sigma_0^1\sigma_0^2+2\sigma_z^1\sigma_z^2\mright]} 
+ \frac{1}{80} 
\sum_{i=0}^{3}\sum_{j=0}^{3}\mleft(\Tr{\mleft[\sigma_z\sigma_i\sigma_z\sigma_i\mright]}\Tr{\mleft[\sigma_0\mright]}+\Tr{\mleft[\sigma_0\mright]}\Tr{\mleft[\sigma_z\sigma_j\sigma_z\sigma_j\mright]} \mright. \notag \\ 
&+& \mleft. \Tr{\mleft[\sigma_z\sigma_i\sigma_0\sigma_i\mright]}\Tr{\mleft[\sigma_0\sigma_j\sigma_z\sigma_j\mright]}+\Tr{\mleft[\sigma_0\sigma_i\sigma_z\sigma_i\mright]}\Tr{\mleft[\sigma_z\sigma_j\sigma_0\sigma_j\mright]}\mright)\notag \\ 
&=& -\frac{2 \times 4}{4+1} + \frac{1}{80}(0+0+0+0) = -\frac{8}{5}.
\eea
Here we used that the trace of $\sigma_z^1\sigma_z^2=(\sigma_z \otimes \sigma_0)(\sigma_0 \otimes \sigma_z)=\sigma_z \otimes \sigma_z$ is zero and in the double sum we use the cancellation from \eqref{cancellation}. 
In a similar way, we obtain for $N$ qubits connected to the same dephasing bath
\bea
\label{deltaFncorrDeph}
\delta F_N\left(\sum_{k=1}^N\sigma_z^k\right)&=& -\frac{1}{d+1} \Tr{\mleft[N \hat{1}_d + \sum_{i\neq j}\sigma_z^i\sigma_z^j\mright]} 
+ \frac{1}{d^2(d+1)} 
\sum_{i=0}^{d^2-1}\sum_{k=1}^N\sum_{l=1}^N\Tr{\mleft[\sigma_z^k\hat{f}_i\sigma_z^l\hat{f}_i\mright]} \notag \\
&=&-\frac{N d + 0}{d+1} + \frac{0}{d^2(d+1)} = -\frac{N d}{d+1} = \sum_{k=1}^N \delta F_N\mleft(\sigma_z^k\mright).
\eea
Here we see that the reduction of the $N$-qubit gate fidelity is the same if all the qubits are subject to uncorrelated dephasing with rate $\Gamma_\phi$ each or correlated dephasing with the same rate. The correlated dephasing acts stronger on some coherences, but weaker on others, such that the combined effect on the average gate fidelity is identical to uncorrelated dephasing. The average gate fidelity is thus not a good indicator for determining whether the dephasing is correlated or not.

For correlated relaxation, we again begin with the case of two qubits. When these qubits are coupled to the same bath, inducing correlated relaxation, the jump operator becomes $\hat{L}=\sigma_-^1+\sigma_-^2$ and we find the fidelity reduction
\bea
\label{deltaF2corrRel}
\delta F_2(\sigma_-^1+\sigma_-^2)&=& -\frac{1}{4+1} \Tr{\mleft[\sigma_+^1\sigma_-^1+\sigma_+^2\sigma_-^2+\sigma_+^1\sigma_-^2+\sigma_+^2\sigma_-^1\mright]} 
+ \frac{1}{80} 
\sum_{i=0}^{3}\sum_{j=0}^{3}\mleft(\Tr{\mleft[\sigma_-\sigma_i\sigma_+\sigma_i\mright]}\Tr{\mleft[\sigma_0\mright]} \mright. \notag \\ 
&+& \mleft. \Tr{\mleft[\sigma_0\mright]}\Tr{\mleft[\sigma_-\sigma_j\sigma_+\sigma_j\mright]}+\Tr{\mleft[\sigma_-\sigma_i\sigma_0\sigma_i\mright]}\Tr{\mleft[\sigma_0\sigma_j\sigma_+\sigma_j\mright]}+\Tr{\mleft[\sigma_0\sigma_i\sigma_+\sigma_i\mright]}\Tr{\mleft[\sigma_-\sigma_j\sigma_0\sigma_j\mright]}\mright) \notag \\
&=& -\frac{2+2+0+0}{4+1} + \frac{1}{80}(0+0+0+0) = -\frac{4}{5}. 
\eea
The corresponding $N$-qubit expression is
\bea
\label{deltaFncorrRel}
\delta F_N\mleft(\sum_{k=1}^N\sigma_-^k\mright)&=& -\frac{1}{d+1} \mleft(N \Tr{\mleft[\sigma_+\sigma_-\mright]} 2^{N-1} + \sum_{i\neq j}\Tr{\mleft[\sigma_+^i\sigma_-^j\mright]} \mright)
+ \frac{1}{d^2(d+1)} 
\sum_{i=0}^{d^2-1}\sum_{k=1}^N\sum_{l=1}^N\Tr{\mleft[ \sigma_-^k\hat{f}_i\sigma_+^l\hat{f}_i\mright]} \notag \\
&=&-\frac{N d + 0}{2(d+1)} + \frac{0}{d^2(d+1)} = -\frac{N d}{2(d+1)} = \sum_{k=1}^N \delta F_N\left(\sigma_-^k\right).
\eea
Together, the results in Eqs.~(\ref{deltaFncorrDeph}) and (\ref{deltaFncorrRel}) give \eqref{eq:dF_Corr_Deph_Rel} in the main text.

Finally, we can also model the correlated noise induced by two-photon relaxation from two qubits with rate $\Gamma_{2p}$ through the jump operator $\hat{L}=\sigma_-^1\sigma_-^2$. Inserting this jump operator in \eqref{deltaFmulti2}, we find
\bea
\label{deltaF2photonRel}
\delta F_2(\sigma_-^1\sigma_-^2)&=& -\frac{\Tr{\mleft[\sigma_+\sigma_-\mright]}\Tr{\mleft[\sigma_+\sigma_-\mright]}}{4+1} 
+ \frac{1}{80} 
\sum_{i=0}^{3}\sum_{j=0}^{3} \Tr{\mleft[\sigma_-\sigma_i\sigma_+\sigma_i\mright]}\Tr{\mleft[\sigma_-\sigma_j\sigma_+\sigma_j\mright]} \notag \\ 
&=& -\frac{1}{4+1} + \frac{0}{80} = -\frac{1}{5}, 
\eea
giving the two-qubit fidelity reduction
\be
\bar{F}=1-\frac{\Gamma_{2p}\tau}{5},
\ee
which is \eqref{eq:dF_TwoPhoton} in the main text.


\end{document}